\documentstyle[prb,aps,floats,epsf]{revtex}  
\begin{document}

\input epsf   

\newcommand{\bi}[5]{Bi$_{#1}$Sr$_{#2}$Ca$_{#3}$Cu$_{#4}$O$_{#5+\delta}$}  
\newcommand{\tc}{T$_{c}$}
\newcommand{\cuo}{CuO\(_{2}\)}
\newcommand{\tstar}{T$^{*}$}
\newcommand{\cm}{cm$^{-1}$}
\newcommand{\aga}{A$_{1g}$} 
\newcommand{\bgb}{B$_{1g}$}
\newcommand{\oxy}[1]{$^{#1}$O}
\newcommand{\dwave}{$d_{{\bf x}^{2}-{\bf y}^{2}}$}
\newcommand{\ybco}[4]{Y$_{#1}$Ba$_{#2}$Cu$_{#3}$O$_{#4-\delta}$}
\newcommand{\lasr}[4]{La$_{#1}$Sr$_{#2}$Cu$_{#3}$O$_{#4-\delta}$}

\twocolumn[
\hsize\textwidth\columnwidth\hsize\csname@twocolumnfalse\endcsname
\draft

\title{On the 590 cm$^{-1}$ B$_{{\bf 1g}}$ feature in underdoped \bi{2}{2}{}{2}{8}}
\author{K. C. Hewitt, N. L. Wang and J. C. Irwin}
\address{Department of Physics, Simon Fraser University\\
Burnaby, B. C. Canada V5A 1S6 \\}
\author{D. M. Pooke}
\address{Industrial Research Limited, P. O. Box 31310\\
Lower Hutt, New Zealand\\}
\author{A. E. Pantoja and H. J. Trodahl}
\address{School of Chemical and Physical Sciences, Victoria University of Wellington \\
PO Box 600, Wellington, New Zealand}
\date{\today}
\maketitle

\begin{abstract}
Raman scattering studies have been performed on underdoped
\bi{2}{2}{}{2}{8}.  
In single crystals underdoped by oxygen removal,
a 590 \cm\ peak is observed in the \bgb\ spectrum.  The feature 
is observed to soften in frequency 
by 3.8\% with isotopic exchange of \oxy{16} by \oxy{18}.  
In contrast, the 590 \cm\ peak is not observed in crystals 
underdoped by Y substitution which suggests that it corresponds
to a disorder induced vibrational mode.  We have also found that 
underdoping leads to a depletion of low energy spectral weight
from regions of the Fermi surface located near the Brillouin zone
axes.

\end{abstract}
\pacs{PACS numbers: 74.25.Gz,74.25.Jb,74.25.Kc,74.62.Dh,74.72.Hs \hspace{1.7in}}
]

One of the interesting electronic properties of underdoped high temperature
cuprate superconductors is the presence of a normal state pseudogap (PG) 
that appears \cite{timu98} at a
temperature \tstar\ which is well above the superconducting transition temperature \tc.  
Evidence for this electronic gap-like feature has been obtained
using many different experimental techniques \cite{timu98}.  In particular,
Angle Resolved Photoemission spectroscopy (ARPES) \cite{mar96,har96,ding96}
and Electronic Raman Scattering (ERS) spectroscopy \cite{chen97a,chen97b,naei98} 
measurements have provided direct evidence of a spectral weight depletion 
from regions of the Fermi surface near ($\pi$,0) in underdoped cuprates.  
Raman scattering 
experiments have also been used \cite{neme97,ope97,quil98,blu97a,blu97} to probe
the energy scale and temperature dependence of the pseudogap.

        Theories of the cuprate pseudogap can be separated into two broad categories 
\cite{maly}.  First there are those \cite{emer95,emer97} that associate the PG with some form of precursor 
superconductivity.  In this case uncorrelated pairs can form at T$>$\tc, with 
phase coherence and hence superconductivity occuring at \tc.  The magnitudes of the PG 
(E$_{g}$) and superconducting gap ($\Delta({\bf k})$) should thus be equal, and furthermore they should 
have the same symmetry.  Thus there should be a smooth evolution from PG to 
superconducting gap (SCG) as 
the temperature is lowered through \tc.  In a second category of theories the loss of 
spectral weight associated with the PG is attributed to short range magnetic correlations 
\cite{whea88,naga90}, or magnetic pairing of some sort \cite{pine97,pine97a}.  In this case the magnitude of the PG 
should be determined by the antiferromagnetic exchange energy J, which in the cuprates is 
much larger than $\Delta({\bf k})$.

Both ERS and ARPES have produced \cite{timu98} similar results for the 
magnitude ($2\Delta_{max}\approx k_{B}T_{c}$) and symmetry (\dwave ) of 
the superconducting gap in optimally doped cuprates.  In underdoped materials,
ARPES measurements \cite{timu98,mar96,har96,ding96} on \bi{2}{2}{}{2}{8}\ (Bi2212) suggest that the
pseudogap evolves continuously into the SCG and thus has magnitude
$E_{g}\approx \Delta$ and \dwave\ symmetry.  The ARPES results are supported by
the Raman measurements of Blumberg et al \cite{blu97a,blu97} who attributed
a weak peak, with frequency $\omega \approx 600 cm^{-1}$ in the \bgb\
spectrum of underdoped Bi2212 to the formation of a bound state, associated
with precursor pairing of quasiparticles above \tc.  However, these results
are in direct contrast to other Raman studies \cite{chen97a,chen97b,naei98,neme97,ope97}
which conclude that the pseudogap does not have \dwave\ symmetry and has a
magnitude $E_{g}\approx J \approx 100 meV > \Delta$, where J is the 
antiferromagnetic exchange constant.

Given the potential significance of the peak near 600 \cm\ we sought to determine whether the
excitation was affected by \oxy{18} substitution.  
We have carried out experiments on two pairs (\tc\ = 51K, 82K) of underdoped samples
of Bi2212, one containing the \oxy{18} isotope and the other \oxy{16}.  While 
confirming the expected shift of previously identified oxygen-related phonons we
have also found that the 590 \cm\ \bgb\ mode shifts with \oxy{18} 
substitution.  This result, combined with experiments on Y-doped Bi2212
have led us to conclude that the 590 \cm\ mode is more appropriately assigned to
an oxygen deficency related vibration.

Through oxygen exchange methods \oxy{16}\ and \oxy{18}\ single 
crystals of Bi2212 were prepared with similar \tc's
in the underdoped regime.  Fully exchanged and underdoped 
\oxy{18}\ samples were prepared by
completing the following steps.  A sample of \oxy{16}-Bi2212 was first 
annealed in vacuum at 500$^{o}$C for four hours.  The ampoule containing
the sample was then evacuated again and backfilled with \oxy{18}\ to a 
pressure of less than one atmosphere.  The sample was then annealed in the \oxy{18}\
atmosphere for twenty-four hours at 850$^{o}$C after which it was gradually
cooled over a period of 15 hours to a temperature of 300$^{o}$C.  Then oxygen
was unloaded by the following procedure.  The ampoule was evacuated, then annealed
at 500$^{o}$C, cooled to 350$^{o}$C and quenched to room temperature.
The \oxy{18}\ samples exhibit a magnetically determined
superconducting onset temperature of 51K.  An unexchanged sample, 
\oxy{16}\ Bi2212, was also prepared using the same final unloading anneal,
and exhibited a superconducting onset temperature of 51K.

The second method of underdoping relies on the use of cation substitution
methods.  In this case the substitution of trivalent yttrium ($Y^{3+}$) for
divalent calcium ($Ca^{2+}$) reduces the hole concentration in Bi2212. 
Adding Y causes \tc\ to fall \cite{wang96}, and yields a non-superconducting 
compound at a Y content of 0.50 \cite{kak96}.  
Single crystals of Y-doped Bi2212 were prepared as described elsewhere \cite{wang96}. 
In the crystals studied Electron Dispersive X-ray (EDX) analysis 
revealed a yttrium content of 0.40 and 0.07, with superconducting transition
temperatures of 30 K and 70 K, respectively.

Raman vibrational spectra in the frequency range 20 - 1000 \cm\
were obtained using the 5145 and 4880 \AA\ lines
of an argon ion (Ar$^{+}$) laser as the excitation source.  The Raman measurements
were carried out in a quasi-backscattering geometry, with the incident laser beam
directed perpendicular to the freshly cleaved ({\bf a},{\bf b}) face of the crystal.

Bi2212 has an orthorhombic structure with the
{\bf a} and {\bf b} axes oriented at 45$^{o}$ to the Cu-O bonds.  
In Raman experiments it is customary to assume a tetragonal structure with
axes parallel to {\bf a} and {\bf b}. 
To facilitate comparison with other cuprates, however, we will consider
a tetragonal cell with {\bf x} and {\bf y} axes parallel to the 
Cu-O bonds.  Another set of axes, {\bf x}' and {\bf y}' are rotated by 
45$^{o}$ with respect to the Cu-O bonds. The 
{\bf x}'{\bf y}' (\bgb ) scattering geometry means that the incident(scattered)
light is polarized along {\bf x}'({\bf y}') and selection of this 
scattering channel enables coupling to excitations having \bgb\ symmetry.
Similarly, the {\bf xy} geometry allows coupling to B$_{2g}$ 
excitations, which transforms as d$_{{\bf xy}}$.  Finally, the diagonal scattering 
geometry {\bf x}'{\bf x}' allows coupling to \aga\ + B$_{2g}$ and {\bf x}{\bf x} to \aga\ + \bgb\
excitations.  Thus, by choosing the polarization of the incident and scattered light one
may select different components of the Raman tensor and thus various
symmetry properties of the excitations.

\begin{figure}[htb]
\centerline{\epsfxsize= 3.5 in \epsffile{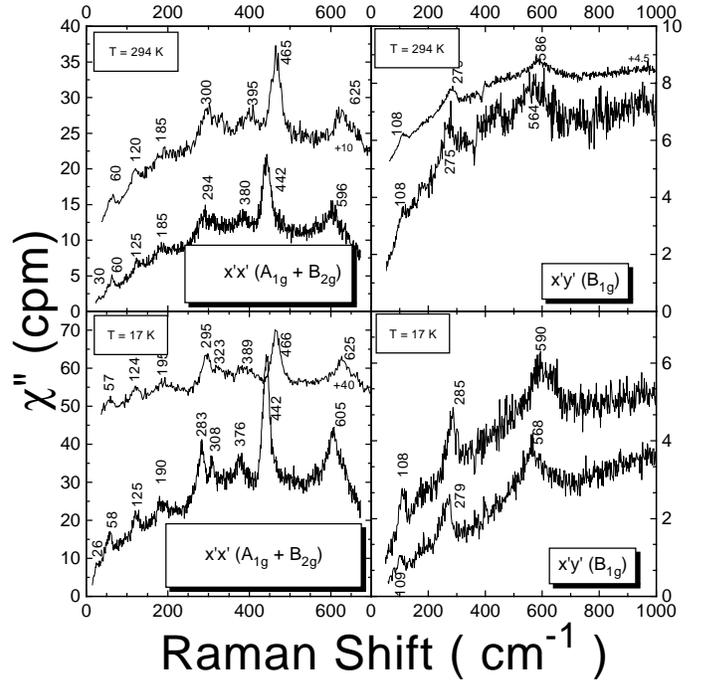}}
\vspace{0.1in}
\caption{The Raman spectra of \oxy{18}\ exchanged (lower curve in each 
panel) and unexchanged \oxy{16}\ (upper curve in each panel) 
Bi2212, taken at 294 and 17 K in the {\bf x}'{\bf x}' (\aga\ + \bgb\ ) and {\bf x}'{\bf y}' ( \bgb\ ) 
geometry.  Both \oxy{18}\ and \oxy{16} crystals have a \tc\ $\approx$ 51K.}
\label{51K-b1g}
\end{figure}

Fig.~\ref{51K-b1g} shows the \bgb\ ({\bf x}'{\bf y}') and (\aga\ + B$_{2g}$) 
({\bf x}'{\bf x}') Raman response functions of the underdoped \oxy{18}\ and 
\oxy{16}\ crystal, taken at room temperature and at T = 17K. 
The observed spectral features may be separated into two categories, 
modes unaffected by \oxy{18}\
substitution and modes softened by \oxy{18}\ substitution.  The low frequency
($\omega < 200 $ \cm) modes are unaffected by \oxy{18}\ substitution
while modes above this frequency show a decrease in frequency.
The heavy metal ion (Bi,Sr) modes have low frequencies and, consistent with
our observation, should be unaffected by \oxy{18}\ substitution. 

In the \bgb\ geometry, the 285 \cm\ mode shows a softening
of -2.1\% while the low frequency mode
at 108 \cm\ shows no shift at all, consistent with the results
of other isotope studies \cite{mar95,pan98}.  The 285 \cm\ mode has been
assigned to out-of-phase vibrations of oxygens in the \cuo\ plane, while
the 108 \cm\ mode has been assigned to b-axis vibrations of Bi \cite{liu92}.

With \oxy{18}\ exchange the 590 \cm\ feature observed in \bgb\ softens 
in frequency by 3.8\%, from 590 to 568 \cm (Fig.~\ref{51K-b1g}).
Also, relative to the continuous background, the 
intensity of the feature increases with decreasing temperature.  We found 
little change in these parameters with varying laser wavelength, at 514.5 nm 
and 488.0 nm. Identical results were obtained for the frequencies and
isotope shift in the second pair (\tc\ = 82K) 
of underdoped crystals. 

\begin{figure}[htb]
\centerline{\epsfxsize= 3.5 in \epsffile{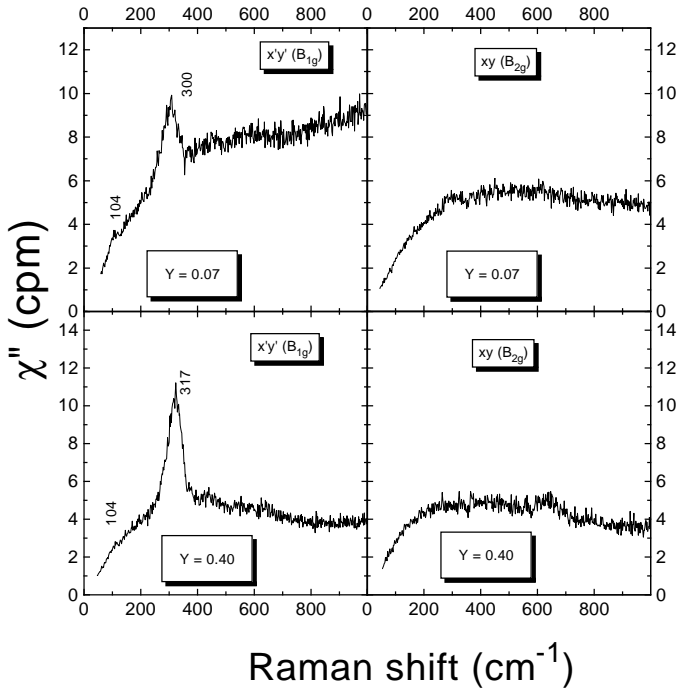}}
\vspace{0.1in}
\caption{The Raman spectra of Y-doped Bi2212 with Y = 0.07 (upper panels)
and Y = 0.40 (lower panels), 
taken at 294K in the {\bf x}'{\bf y}' (B$_{1g}$) and {\bf xy} (B$_{2g}$) geometry.}
\label{ydopedb12g}
\end{figure}

Since the 590 \cm\ \bgb\ feature is only observed in underdoped crystals, and since
the mass dependence is consistent with an oxygen related vibration,
we sought to determine whether underdoping without a corresponding
reduction in $\delta$ affects the properties of the mode.  This was achieved 
by substitution of Y$^{3+}$ for Ca$^{2+}$.  As shown in Fig.~\ref{ydopedb12g}
, for a [Y] = 0.4 and 0.07, the
590 \cm\ mode is completely {\it absent} from the \bgb\ spectra, as found by 
Kendziora \cite{ken97} for a another underdoped crystal ([Y] = 0.15).  
This result
implies that the 590 \cm\ \bgb\ mode is associated with oxygen
removal, and is thus expected to be related to an oxygen deficency type of
vibration.

To speculate on the origin of the 590 \cm\ mode one can first note that there 
appears to be an analogous situation in Y123.  In Y123, a mode appears 
\cite{hadj91,iva95} 
in the \bgb\ channel at about 600 \cm\ when the oxygen concentration is 
reduced below its optimum value of x $\approx$ 6.95.  In this case 
underdoping is associated with oxygen removal from the chains and the 
creation of chain fragments.  Thus many oxygen atoms in the chains are no 
longer situated at centers of inversion and previously forbidden modes can 
become Raman active.  In the case of Bi2212 one can note that the Bi-O layers 
are weakly bonded to one another and oxygen intercalation or exchange takes 
place in these planes\cite{pan98,john95}.  There is also an incommensurate structural modulation 
along the b-axis in the Bi-O planes \cite{john95} that results from oxygen 
non-stoichiometry and a lattice mismatch between the Bi-O rocksalt layers 
and the perovskite layers.  The removal of oxygen from Bi2212 should thus 
give rise to the formation of buckled Bi-O modulation chain fragments, and 
hence to disorder that would lead to the activation of normally Raman 
forbidden modes.  Thus one might speculate that the 590 \cm\ mode in Bi2212 
results from b-axis stretching vibrations of oxygen atoms in the Bi-O 
layer.  The buckling associated with the modulation could lead to a broadening 
of the mode, consistent with observation.

        In previous experiments on Y123 \cite{chen97b} and La214 \cite{naei98} 
it was found that underdoping led to a reduction of the scattering intensity 
in the \bgb\ channel.  In other words spectral weight was lost from regions of 
the Fermi surface located  
near ($\pi$,0) and symmetry related points.  The present 
spectra suggest that a similar depletion occurs in underdoped Bi2212.

\begin{figure}[htb]
\centerline{\epsfxsize= 3.5 in \epsffile{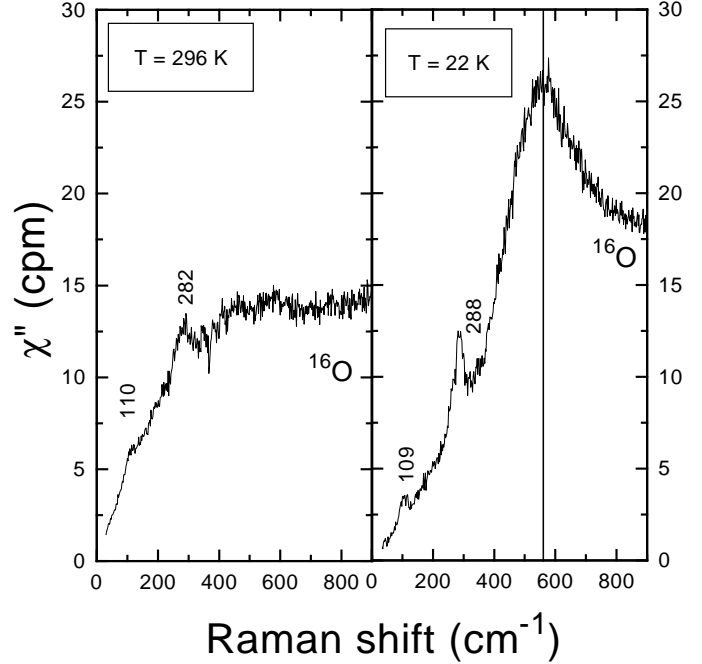}}
\vspace{0.1in}
\caption{The Raman spectra of optimally doped \oxy{16} Bi2212 (T$_{c} $ = 90K), 
taken at 296 and 22 K in the and {\bf x}'{\bf y}' (B$_{1g}$) geometry.}
\label{optimalb1g}
\end{figure}

In the optimally 
doped material (Fig.~\ref{optimalb1g}) the continuum in the \bgb\ spectrum is 
characterized by a count rate of approximately 15 cpm (T=296K) and a strong 
superconductivity induced renormalization is observed 
when the sample is cooled through \tc.  

In the underdoped compound with 
\tc\ = 51K the strength of the \bgb\ spectrum is reduced to about 4 cpm
(Fig.~\ref{51K-b1g}).  In the 
Y substituted compounds (Fig.~\ref{ydopedb12g}), for the 
sample with [Y] = .07 (\tc\ $\approx$ 70K),  $\chi"(B_{1g}) \approx 9 cpm $ 
and for [Y] = 0.4 (\tc\ $ \approx$ 30K) one has $\chi"(B_{1g})\approx 4 cpm $.  
One can also note that there is no \bgb\ renormalization at \tc\ for the 
underdoped compounds.

\begin{figure}[htb]
\centerline{\epsfxsize= 3.5 in \epsffile{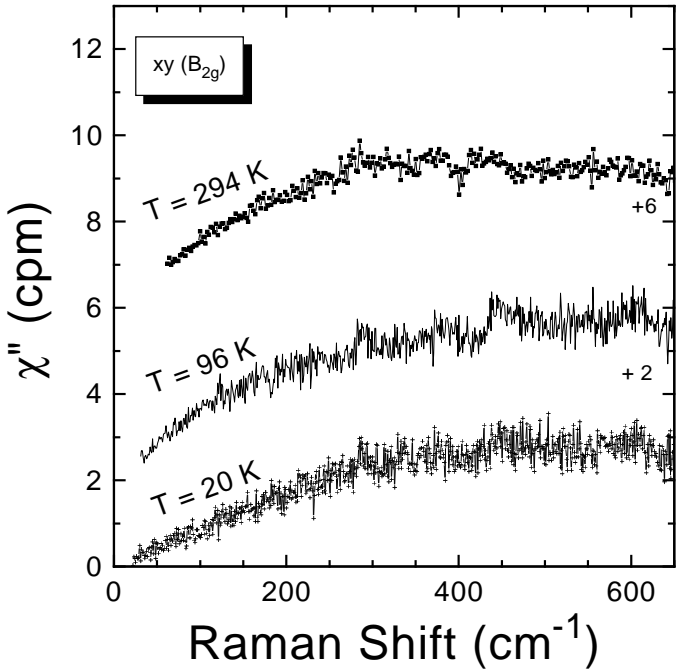}}
\vspace{0.1in}
\caption{The Raman spectra of \oxy{18}\ exchanged Bi2212 (\tc\ = 51K), taken 
at three temperatures (294, 96 and 20 K) in the {\bf x}{\bf y} ($B_{2g}$) 
scattering geometry.}  
\label{51K-b2g}
\end{figure}

In the $B_{2g}$ channel the strength of the spectrum (Fig.~\ref{ydopedb12g}) 
is approximately the 
same for both Y doped compounds and is given by $\chi"(B_{2g}) \approx 4cpm$.  
In the \oxy{18}-Bi2212 crystal the $B_{2g}$ spectrum (Fig~\ref{51K-b2g}) also 
has $\chi"(B_{2g}) \approx 4cpm$. It is interesting to note that there does
not appear to be a well-defined renormalization of the B$_{2g}$ spectrum at
\tc.

We have determined the ratio $R = \chi"(B_{1g})/\chi"(B_{2g})$ for the above
compounds and find that it decreases by about a factor of four as the doping
level is decreased from optimum (\tc = 90K) to the lowest value examined
here (\tc = 40K).  A corresponding hole concentration can be estimated \cite{tall95} 
from, $\frac{T_{c}}{T_{c,max}} = 1 - 82.6(p-p_{opt})^2$. The
magnitude and rate of decrease in R is similar to the Bi2212 results of
Opel et al\cite{ope97} and to values found for Y123 \cite{chen97b} and La214
\cite{naei98}.  Since $\chi"(B_{2g})$ is approximately independant of doping,
the decrease in R is attributed to a depletion of spectral weight in the \bgb\
channel.

Since the decrease in strength of the Bi2212 \bgb\ spectrum with decreasing
doping is similar to that observed in Y123 and La214, 
we suggest that
the depletion of \bgb\ spectral weight in underdoped compounds is a 
characteristic feature of the cuprates.
The depletion is attributed \cite{chen97b,naei98} to the existence of 
a pseudogap of magnitude E$_{g} \approx J$ that is 
localized near ($\pi$,0).  As the doping level is decreased more and more of 
the Fermi surface becomes depleted.  Or, from another perspective, the Fermi 
surface in underdoped compounds appears to consist of arcs centered near the 
diagonal directions in k-space.  Our results thus suggest that the lengths of 
the arcs decreases with 
decreasing doping \cite{naei98}.  One can also see (Fig.~\ref{51K-b1g}) 
that, at low energies, $\chi"(B_{1g})$ decreases with temperature, as 
observed \cite{quil98} previously in Bi2212.

In conclusion, the 590 \cm\ \bgb\ mode is found in the Raman  
spectra of Bi2212 in which underdoping is achieved by the
process of oxygen depletion.  This mode is absent in spectra obtained from crystals
in which underdoping is achieved by cation substitution.  The mode
softens by -3.8\% with \oxy{18}\ substitution. Based on these observations 
we conclude that the 590 \cm\ feature is vibrational in origin and not
associated with pseudogap formation.  However, in this regard, underdoping
of Bi2212 leads to a depletion of spectral weight from regions of the Fermi 
surface near ($\pi$,0).  The strength of this depletion is consistent with
that observed in other cuprates.

We would like to thank the Natural Sciences and Engineering Research Council
of Canada for financial support, and T. P. Devereaux, 
R. G. Buckley and J. L. Tallon for useful discussions.

\bibliographystyle{prsty}

\end{document}